\documentclass[prb, twocolumn, showpacs]{revtex4}
\usepackage{amssymb}
\usepackage{amsmath}
\usepackage{graphicx}
\usepackage{color}
\makeatletter

\begin{document}
   \DeclareGraphicsExtensions{.pdf,.gif,.jpg.eps}
\title{Diagrammatic Quantum Monte Carlo solution of the two-dimensional Cooperon-Fermion model}
\author{Kai-Yu Yang$^{1,2}$, E. Kozik$^{1}$, Xin Wang$^{3}$, M. Troyer$^{1}$ }
\address{
${1}$ Theoretische Physik, ETH Zurich, CH-8093 Zurich \\
${2}$ Department of Physics, Boston College, Chestnut Hill, Massachusetts 02467, USA\\
${3}$ Condensed Matter Theory Center, Department of Physics, University of Maryland, College Park, MD 20742
}
\date{\today}

\begin{abstract}
We investigate the two-dimensional cooperon-fermion model in the correlated regime with a new continuous-time diagrammatic determinant quantum Monte Carlo (DDQMC) algorithm. We estimate the transition temperature $T_{c}$, examine the effectively reduced band gap and cooperon mass, and find that delocalization of the cooperons enhances the diamagnetism. When applied to diamagnetism of the pseudogap phase in high-$T_{c}$ cuprates, we obtain results in a qualitative agreement with recent torque magnetization measurements. 
\end{abstract}

\pacs{05.30.Fk,  74.20.Fg,  05.10.Ln, 05.70.Fh}

\maketitle

\section{Introduction}
The cooperon-fermion model (for a review, see Ref. \onlinecite{Chen-PR-05}) is a basic model for superconductivity that has widely been adopted to explain the BCS-BEC crossover in ultra-cold fermionic atomic gases \cite{Chen-PR-05, Holland-PRL-01, Ohashi-PRL-02} and high-$T_c$ superconductivity.\cite{Chen-PR-05, Minca-RMP-90, Friedberg-PRB-1989, Domanski-arxiv-03, Dima-97, Micnas-PRB-07, Ranning-PC-95,Neto-PRB-01, Perali-PRB-00, Altman-PRB-02, Tsai-PRB-06, Alexandrov-04}
The resonantly paired fermions, or cooperons, in this model can either be locally bound pairs of small polarons due to extremely strong electron-phonon coupling,\cite{Robaszkiewicz-PRB-87} or localized Cooper pairs due to strong local pairing as might be the case in high-$T_c$ superconductors \cite{Altman-PRB-02}, or molecular bosons in ultra-cold atoms.\cite{Chen-PR-05, Holland-PRL-01, Ohashi-PRL-02} The potential existence of finite energy cooperons with a local attraction has also been put forward a few years ago in a simple semiconductor system.\cite{Nozieres-EPJB-99} 
One recent work based on the four-leg ladder Hubbard model \cite{KRT10} as well as earlier studies in the quasi-2D ladder Hubbard model \cite{LeHur-Review} reveals the important role of the cooperon excitations in the transition from insulating state to superconducting state. 

In cuprates, the interplay between the finite energy cooperon excitations around the antinode and the truncated Fermi surface around the node has recently been proposed as a possible mechanism for the superconductivity in a two-gap scenario.\cite{YRZ, RYZ-Review}
The dominant underlying mechanism for driving superconductivity in this model is the scattering two electrons involving a virtual cooperon $c_{\uparrow, \boldsymbol{k}} + c_{\downarrow, -\boldsymbol{k}} \xrightarrow{b_{\boldsymbol{k}=0}}  c_{\uparrow, \boldsymbol{k}'} + c_{\downarrow, -\boldsymbol{k}'}$. 
Compared with the attractive Hubbard model, the cooperon-fermion model has much richer physics since it has the complete dynamical information of the interaction between two fermions and the delocalization of cooperons with decreasing temperature.

So far most work on this model is done at the mean field level, with either the T-matrix method or various other methods going beyond simple mean field theory by including more diagrams \cite{Chen-PR-05} but there have been few unbiased calculations.
A recent exact diagonalization study of this model has been limited by small sizes and special geometries. \cite{ED-BF} A direct quantum Monte Carlo simulation usually suffers from a sign problem,\cite{sign-problem} except for certain models and algorithms, such as determinant quantum Monte Carlo simulations of the attractive Hubbard model.

In this paper, we simulate the two-dimensional cooperon-fermion model using a continuous-time diagrammatic determinant quantum Monte Carlo method (DDQMC). 
The Hamiltonian has the form:
\begin{equation}
H=   \sum_{\sigma,\boldsymbol{k}} (c^{\dag}_{\boldsymbol{k},\sigma}\epsilon^{f}_{\boldsymbol{k}} c_{\boldsymbol{k},\sigma} +b^{\dag}_{\boldsymbol{k}}\epsilon^{b}_{\boldsymbol{k}} b_{\boldsymbol{k}})
 + U\sum_{i}   (  c^{\dag}_{i, \uparrow}c^{\dag}_{i,\downarrow}b_{i} + h.c.  )
\label{eq:H}
\end{equation}
where $c^{\dag}_{\sigma} (c_{\sigma})$ is the fermionic creation (annihilation) operator with spin $\sigma = \{\uparrow, \downarrow \}$ and $b^{\dag} (b)$ is the bosonic creation (annihilation) operator of a cooperon. 
The interaction term  $U$ leads to the $s$-wave pairing of fermions mediated by the originally localized cooperons with a band gap $\Delta$ and the delocalization of cooperons at low temperatures. 
The bare dispersions are $\epsilon_{\boldsymbol{k}}^{f}$ = $2t_{f}[2- \cos(k_{x}) - \cos(k_{y}) ] + \mu$, and $\epsilon_{k}^{b} = 2t_{b}(2- \cos(k_{x}) - \cos(k_{y}))  + \Delta$. It is a trivial generalization to include attractive interactions between the fermions.

By integrating out the bosonic degrees of freedom, an effective action for the fermionic part can be obtained, which has a form similar to the attractive Hubbard model but with full dynamic properties:
\begin{equation}
\begin{split}
&S^{\rm eff}_{f}(\bar{\psi}_{\sigma}, \psi_{\sigma}) = \int _{0}^{\beta} d\tau \sum_{\boldsymbol{k},\sigma} \bar{\psi}_{\boldsymbol{k}, \sigma} (\partial_{\tau} + \epsilon^{f}_{\boldsymbol{k}}) \psi_{\boldsymbol{k},\sigma} \\ 
&\!\!+ U^{2} \int_{0}^{\beta}  \int_{0}^{\beta}  d\tau d\tau ^{\prime} \sum_{i, i^{\prime}} \bar{\psi}_{i,\tau,\uparrow} \bar{\psi}_{i,\tau,\downarrow} G^{b,0}_{r_{i}-r_{i\prime},\tau-\tau^{\prime}}  {\psi}_{i^{\prime},\tau^{\prime},\downarrow} {\psi}_{i^{\prime},\tau^{\prime},\uparrow}
\end{split}
\end{equation}
where $G^{b,0}_{r_{i} -r_{i\prime},\tau-\tau^{\prime}}$ is the bare bosonic Green's function.
The attractive Hubbard model can be obtained by approximating $U^{2} G^{b,0}_{r_{i}-r_{i\prime},\tau-\tau^{\prime}}  =U_{H} \delta_{i, i\prime} \delta_{\tau, \tau\prime}$ with  $U_{H} \sim -U^{2}/\Delta_{\rm eff}$ ($\Delta_{\rm eff}$ is the renormalized band gap of the cooperons) including the dominant contribution from the renormalized $\boldsymbol{k}=0$ cooperon. 

We develop a continuous-time DDQMC algorithm, similar to the algorithms for an attractive Hubbard model\cite{Burovski-NJP-06, Burovski-PRL-08} for the cooperon-fermion model, and obtain the numerically exact solution to the model at the filling value $n=0.12$. The Kosterlitz-Thouless (KT) transition temperature is estimated from the finite-size scaling of the fermion pair correlation function and the cooperon Green's function. The renormalization of the cooperon band characterized by the effective gap $\Delta_{\rm eff}$ and the effective mass $m_{\rm eff}$ are examined carefully. 
 Applying these results to study the strong diamagnetism recently observed in cuprates, we find that the renormalization of the cooperon band will enhance the diamagnetism dramatically at low temperatures, which qualitatively agrees with the experimental data. \cite{Lu-epl}

\section{The algorithm}

\begin{figure}[tbf]
\centerline
{
\includegraphics[width = 6.0cm, height =5.0cm, angle= 0] 
{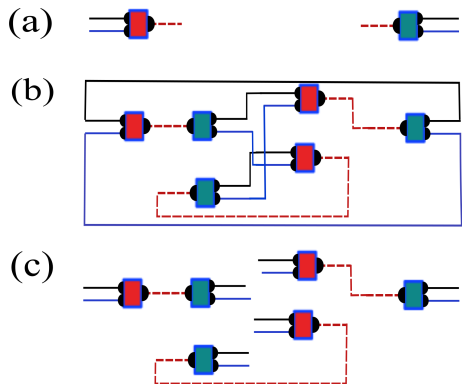} 
}
 \caption[]
 {(Color online.) (a): elementary interaction events (vertices): two fermions are combined onto a cooperon (left) and vice versa (right). (b): a diagram contributing to the partition function $Z$.
 (c): a diagram sampled in the DDQMC algorithm: open ends of the fermionic lines imply a sum over all the ways of connecting the vertices by the fermionic lines, which is represented by the corresponding determinant. }
\label{fig:event}
\end{figure}

Continuous-time DDQMC algorithms have been applied successfully in the past to obtain the critical temperature in the BCS-BEC crossover\cite{Burovski-NJP-06, Burovski-PRL-08}
The sign problem in these algorithms is avoided by collecting all Feynman diagrams with the same distribution of vertices as a single configuration which turns out to have a positive-definitive weight. 
This approach can also be applied to the cooperon-fermion model eliminating the severe sign problem coming from the permutations of the fermionic lines. 
In the interaction picture, the partition function for the model  (\ref{eq:H}) can be expressed as
\begin{equation}
\begin{split}
Z &= 
  {\rm Tr} \Bigl[e^{-\beta (H_{f,\uparrow}^{0} + H_{f,\downarrow}^{0})} e^{-\beta H_{b}^{0}}  \sum_{n} \Big[\int_{0}^{\beta} \int \Big]^{n}  \\
  &\cdot\frac{1}{n!} \prod_{i=1,n} dr_{i} d\tau_{i} \mathcal{T}_{\tau}( -U 
c^{\dag}_{r_{i},\tau_{i},\uparrow}c^{\dag}_{r_{i},\tau_{i},\downarrow}b_{r_{i},\tau_{i}} + h.c.)\Bigr]
\end{split} \label{EQ:Z} 
\end{equation}
where $H_{f, \sigma}^{0}$ in the Hamiltonian of free fermions with the spin $\sigma$ and $H_{b}^{0}$ is the Hamiltonian for free cooperons, the bilinear in $c^{\dag}_{\boldsymbol{k},\sigma}$, $c_{\boldsymbol{k},\sigma}$ and $b^{\dag}_{\boldsymbol{k}}$, $b_{\boldsymbol{k}}$ terms respectively in Eq.~\eqref{eq:H}. To describe the Feynman diagrams, we define two different kinds of events, shown in Fig.~\ref{fig:event}(a), representing the process of combining two fermions with opposite spins into one cooperon and the reverse process. For a typical Feynman diagram like the one shown in Fig.~\ref{fig:event}(b), the vertices are connected by the bare single particle propagators
\begin{align}
G^{f,0}_{r^{a}_{i}-r^{c}_{j}, \tau^{a}_{i}-\tau^{c}_{j}} &= - {\rm Tr}\Bigl[ e^{-\beta H_{f,\sigma}^{0}} \mathcal{T}_{\tau}c_{r^{a}_{i},\tau^{a}_{i},\sigma}  c^{\dag}_{r^{c}_{j},\tau^{c}_{j},\sigma} \Bigr]  \\
G^{b,0}_{r^{c}_{i}-r^{a}_{j}, \tau^{c}_{i}-\tau^{a}_{j}} &= - {\rm Tr}\Bigl[ e^{-\beta H_{b}^{0}} b_{r^{c}_{i},\tau^{c}_{i} }  b^{\dag}_{r^{a}_{j},\tau^{a}_{j}} \Bigr]
\end{align}
The superscript $a/c$ on $r_{i,j}$ are for the events with annihilation and creation of a pair of fermions (in accompany of the creation and annihilation of a cooperon), respectively.
By applying the Wick's theorem, the partition function can be rewritten as
\begin{equation}
\begin{split}
Z =& \sum_{n}  U^{2n} \Big[ \int  _{0}^{\beta} \int \Big]^{2n}\frac{1}{n!n!} \Bigl[ \prod_{i=1,n} d^{D}r^{c}_{i}d^{D}r^{a}_{i} d\tau^{c}_{i}d\tau^{a}_{i} \Bigr] \\
& \cdot\det A_{{S}_{n}^{'},\uparrow} \det A_{{S}_{n}^{'},\downarrow} \text{Perm}(B_{{S}_{n}^{'}})
\end{split}
\end{equation}
where $S_{n}^{'}$ represents the configuration including all possible ways of connecting a specific distribution of vertices with the propagator lines. The matrix components $[A_{S_{n}^{'},\sigma}]^{i,j} =  G^{f,0}_{r^{a}_{i}-r^{c}_{j}, \tau^{a}_{i}-\tau^{c}_{j}}$, and $[B_{S_{n}^{'}}]^{i,j} = - G^{b,0}_{r^{c}_{i}-r^{a}_{j}, \tau^{c}_{i}-\tau^{a}_{j}}$. 
The determinant of the matrix $A$ comes from the anti-commutation relation of fermions, while the permanent Perm$(B_{S_{n}^{'}})$ originates in the commutation relation between cooperons. In contrast to determinants, which can be efficiently evaluated, the calculation of a permanent is an exponentially hard problem. Thus, we evaluate the permanent by individually sampling all permutations of the bosonic lines. A 
 typical diagram $\widetilde{S}_{n}$ encountered in the Monte Carlo sampling is shown in Fig.~\ref{fig:event}(c). The open ends of fermionic lines indicate that all possible connection ways of the fermion lines are summed up in the determinant, while the connection between the cooperon lines is fixed, indicating that the specific connections are sampled individually. Summing the fermion lines into a determinant completely removes the fermionic sign problem. However, sampling the permanent gives us a small remaining sign problem, which is tractable since the distribution of  values in $B_{S_{n}^{'}}$ is  dominantly positive.
 
 The weight of a configuration  $\widetilde{S}_{n}$ is
\begin{equation}
Z_{\widetilde{S}_{n}} =\frac{U^{2n} } {( L^{2} \beta )^{2n}} \det A_{\widetilde{S}_{n},\uparrow} \det A_{\widetilde{S}_{n},\downarrow} \prod_{i=1,n} B^{i,\mathcal{P}_{i}}_{\widetilde{S}_{n}},
 \label{eq:Z}
\end{equation}
where $L^{2}$ is the spatial size of the system with periodic boundary conditions. Thermal averages are calculated by sampling all possible configurations $\widetilde{S}_{n}$.

 \begin{figure}[tbh]
\centerline
{
\includegraphics[width = 7.5cm, height =4.0cm, angle= 0] 
{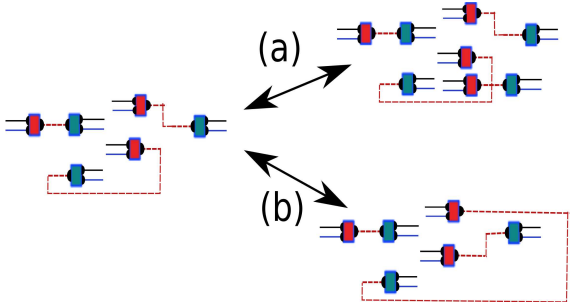} 
}
 \caption[]
 {(Color online.) Two sets of complementary Monte Carlo updates: (a) adding or removing one pair of vertices, and (b) swapping the end points of cooperon lines.}
\label{fig:update}
\end{figure}

Our algorithm implements three different Monte Carlo updates: creating or deleting one pair of vertices which changes the order from $n$ to $n\pm1$, and swapping the connection of cooperon lines  as shown in Fig.~\ref{fig:update}. In order to improve the efficiency of sampling we pick a pair of times with a probability proportional to the cooperon's bare Green's function $G^{b,0}_{r,\tau}$.
Adding a pair of vertices to go from configuration $\widetilde{S}_{n}$ to $ \widetilde{S}_{n+1}$ is accepted with an acceptance ratio of 
\begin{align}
R_{\rm add} = \min \bigg( 1,&U^{2}  \frac{ \left| \det A_{\widetilde{S}_{n+1},\uparrow} \right| ^{2}} {\left| \det A_{\widetilde{S}_{n},\uparrow} \right | ^{2}} \left| B_{\widetilde{S}_{n+1}}^{n+1,\mathcal{P}_{n+1}} \right|  \nonumber \\ 
& \frac{\beta L^{2}} {n+1} \frac{\int_{0}^{\beta} d\tau \sum_{r} G^{b,0}_{r,\tau}} {G^{b,0}_{r,\tau}}
\bigg),
\label{eq:R}
\end{align}
and a corresponding equation for the removal.

For the self-complementary process of swapping the connection of cooperon Greens functions
the acceptance ratio is
\begin{equation}
R_{\rm flip} =\min \left(1, {\left| \prod_{i} B^{i,\mathcal{P}(i)}_{\widetilde{S}_{\mathcal{P}(n)}}  \right|} \Big/ {\left|  \prod_{i} B^{i,\mathcal{P'}(i)}_{\widetilde{S}_{\mathcal{P'}(n)}}   \right|} \right)
\end{equation}

The fermionic and cooperon Green's functions can be measured as
\begin{align}
G^{f}_{R-R^{\prime}, \tau-\tau^{\prime},\sigma} &= \left\langle{\det \widetilde{A}_{\widetilde{S}_{n},\sigma}} / {\det {A}_{\widetilde{S}_{n},\sigma}} 
\right\rangle _{\rm  MC} \label{eq:Gf}\\
G^{b}_{R-R^{\prime}, \tau-\tau^{\prime}} &= \Bigl\langle  G^{b,0}_{R-R^{\prime}, \tau-\tau^{\prime}}  \label{eq:Gb} \\
&+\sum_{l} \frac{G^{b,0}_{R-r^{a}_{l}, \tau - \tau^{a}_{l}} G^{b,0}_{r^{c}_{l}-R^{\prime}, \tau^{c}_{l}-\tau^{\prime}} }{G^{b,0}_{r^{c}_{l}-r^{a}_{l},\tau^{c}_{l}-\tau^{c}_{l}}} \Bigr\rangle_{\rm MC} \notag 
\end{align}
The matrix $\widetilde{A}_{\widetilde{S}_{n}, \sigma}$ is an $(n+1) \times (n+1)$ matrix extending $A$ by  an extra column and row corresponding to the open vertices $ c_{(R^{\prime},\tau^{\prime}) }^{\dag}$ and $ c_{(R, \tau)}$. The notation $\langle .. \rangle_{\rm MC}$ denotes the Monte Carlo average. The particle-particle correlation function of fermions $\langle \mathcal{T}_{\tau} c_{R,\tau,\downarrow}c_{R,\tau,\uparrow} c^{\dag}_{R^{\prime}, \tau^{\prime}, \uparrow}c^{\dag}_{R^{\prime}, \tau^{\prime}, \downarrow}  \rangle$ is measured as
\begin{equation}
\chi^{\rm pp}_{R-R^{\prime}, \tau-\tau^{\prime}} = \left\langle \left| \frac{\det \widetilde{A}_{\widetilde{S}_{n},\sigma}}{\det {A}_{\widetilde{S}_{n},\sigma}} \right|^{2} \right\rangle _{\rm MC} 
\end{equation}
with the double occupancy characterizing the local pairing strength $\langle n_{\uparrow} n_{\downarrow} \rangle $
=$\chi^{\rm pp}_{R=(0,0), \tau=0^{-}}$. 
The vertex correlation contribution to the particle-particle correlation of fermions, which indicates the formation of coherent cooper pairs, is defined as
\begin{equation}
\begin{split}
\chi^{\rm od}_{R-R^{\prime}, \tau-\tau^{\prime}} &=\left\langle \mathcal{T}_{\tau} c_{R,\tau,\downarrow}c_{R,\tau,\uparrow} c^{\dag}_{R^{\prime}, \tau^{\prime}, \uparrow}c^{\dag}_{R^{\prime}, \tau^{\prime}, \downarrow} \right \rangle  \\ 
&- \left\langle \mathcal{T}_{\tau} c_{R,\tau,\uparrow} c^{\dag}_{R^{\prime}, \tau^{\prime}, \uparrow} \right\rangle \left  \langle \mathcal{T}_{\tau}c_{R,\tau,\downarrow}c^{\dag}_{R^{\prime}, \tau^{\prime}, \downarrow} \right\rangle 
\label{eq:chi-od}
\end{split}
\end{equation}
The cooperon Green's function and $\chi^{od}$ in the long wave-length and static limit for our finite system with periodic boundary conditions can be defined as $G^{b}_{\boldsymbol{k}=0,\omega =0} = \int_{0}^{\beta} d\tau \sum_{r} G^{b}_{r,\tau}$, and $\chi^{od}_{\boldsymbol{k}=0,\omega =0} = \int_{0}^{\beta} d\tau  \sum_{r} \chi^{od}_{r,\tau}$ with $r$ being the distance confined by system size.

 \begin{figure}[tb]
\centerline
{
\includegraphics[width = 7.5cm, height =2.0cm, angle= 0] 
{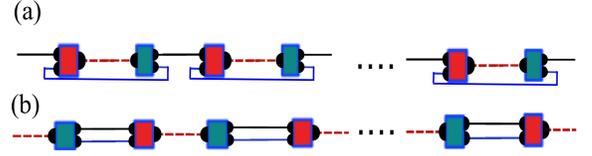} 
}
 \caption[]
 {(Color online.) The Feynman diagrams comprising the fermionic (a) and cooperon (b) Green's function within the random phase approximation (RPA) .}
\label{fig:RPA}
\end{figure}

In addition to the DDQMC results, we also show results of random phase approximation (RPA) calculations which only take into account the contribution from a truncated set of Feynman diagrams,  whose diagrammatic representations are shown in Fig.~\ref{fig:RPA}. The cooperon and fermionic Green's functions have the general form
\begin{align}
[G^{f/b}_{\boldsymbol{k},\omega}]^{-1} 
= [G^{f/b,0}_{\boldsymbol{k},\omega}]^{-1} 
- \Sigma^{f/b}_{\boldsymbol{k},\omega}
\end{align}
 with the self energies $\Sigma^{f/b}_{\boldsymbol{k},\omega}$ estimated by only including ladder diagrams.
 RPA works well at weak coupling and high temperature region, and is useful as a test for our numerical solution in that limit.

\section{Main results}

For our simulations we choose $U=1$ as the unit of energy and set $t^{f} = 1$, $t^{b} = 0.5$. The bare cooperon band lies above the bottom of the fermionic band with the offset $\Delta=0.75$ and the total charge density is fixed at $n^{f}_{\uparrow} + n^{f}_{\downarrow}+ 2n_{b}=0.12\pm0.002$.
In the parameter regime we are interested in, the chemical potential $\mu$ is around $0.22-0.45$, resulting in the effective renormalized gap $\Delta_{\rm eff} \le 0.3$ and the corresponding effective attractive Hubbard interaction $|U_{H}| \sim U^{2}/\Delta_{\rm eff} > 3$ at $\beta > 3$. For the finite-size scaling analysis, we use the set of linear system sizes $L = 11, 15, 21, 25$. The expectation values and the error bars are obtained from 96 independent sampling processes with different random number seeds. Each measurement is made after 2-3 autocorrelation times, {\it i.e.} around one measurement per 100$L^{2}$ Monte Carlo steps at high temperatures ($\beta \sim 3$), and 3000-5000 $L^{2}$ steps for low temperatures ($\beta \sim17$).

\begin{figure}[tb]
\centerline
{
\includegraphics[width = 5.5cm, height =9.0cm, angle= 270] 
{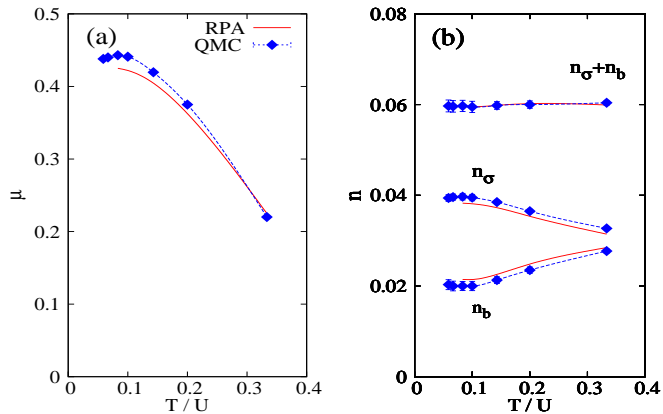}
}
 \caption[]
 {(Color online.) (a): the chemical potential versus temperature $T/U$. 
 (b): the particle density $n_{\sigma}$ of a single spin component and the cooperon density $n_{b}$ versus temperature $T/U$. DDQMC results in thermodynamic limit (blue points) are extrapolated to the infinite system size.}
\label{fig:mu}
\end{figure}

Fig.~\ref{fig:mu}(a) shows the temperature dependence of the chemical potential $\mu$. The solid curve is the RPA result with a mean-field critical temperature $T_{c}^{MF} \sim 0.09$ characterized by the closure of the effective gap $\Delta_{\rm eff}$. 
Blue diamonds are the chemical potential at thermodynamic limit determined by DDQMC. 
RPA generally overestimates the interplay between fermions and cooperons at low temperatures due to the logarithmic divergence leading to a larger value of the cooperon's self-energy from the simple particle-particle bubble diagram. 
As a consequence, the RPA chemical potential is lower than the exact value in this regime.
However, in the high-temperature limit, the RPA calculation provides a consistency check for DDQMC and there we observe a perfect agreement.
Fig.~\ref{fig:mu}(b) shows the particle density $n_{\sigma}$ and $n_{b}$ at thermodynamic limit. 
The cease of the suppression of $n_{b}$ as the temperature is decreased, i.e. the flattening out of $n_{b}$ v.s. $T$ at low $T(<0.1)$, indicates an approach to the KT transition. 

\begin{figure}[tbh]
\centerline
{
\includegraphics[width = 5.5cm, height =9.0cm, angle= 270] 
{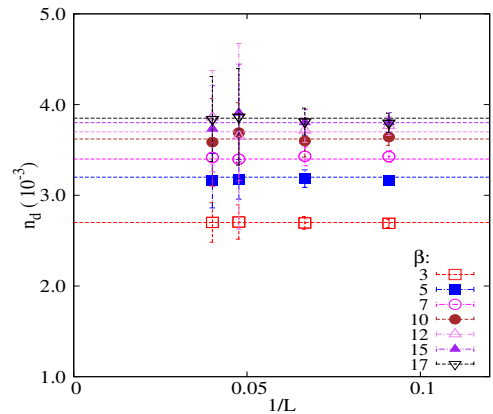}
}
 \caption[]
 {(Color online.) The system size dependence of the onsite double occupancy $n_{d}$ for various values of $\beta$. }
\label{fig:double-n}
\end{figure}

As seen from Fig.~\ref{fig:double-n}, the double occupancy $n_{d} =\langle n_{i,\uparrow}n_{i, \downarrow} \rangle  \gg \langle n_{i,\uparrow} \rangle \langle n_{i, \downarrow} \rangle$ reveals a strong on-site pairing mediated by cooperons. 
The behavior of $n_{d}$ is determined by two aspects: 
(i) the competition between the potential energy gain from pairing and the corresponding kinetic energy loss, and (ii) the balance between the particle number of fermions and cooperons. However, the latter is not expected to play an important role at low temperatures due to the plateau in $n_{b}(T)$.
The monotonous increase of $n_{d}$ with lowering the temperature is analogous to recent DMFT results for the attractive Hubbard model. \cite{Schollwock-prl-01}
We also note that a different low-$T$ behavior of $n_{d}$ for the attractive Hubbard model has been reported in early QMC studies. \cite{Singer-PRB-96}

\begin{figure}[b]
\centerline
{
\includegraphics[width = 5.5cm, height =9.0cm, angle= 270] 
{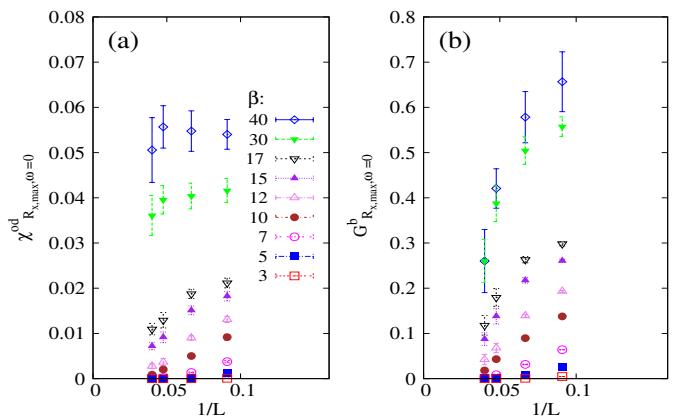}
}
 \caption[]
{(Color online.) (a): system-size dependence of off-diagonal order $\chi^{\rm od}_{R_{x,{\rm max}},\omega =0}$ at distance $R_{x,{\rm max}}$. (b): the cooperon Green's function $G^{b}_{R_{x,{\rm max}}, \omega =0}$. $R_{x,{\rm max}} = (L-1)/2$ is the maximal distance along the $x$ direction in our system with periodic boundary conditions.
}
\label{fig:scaling}
\end{figure}
The 2D superconducting state is characterized by algebraically decaying off-diagonal order \cite{CNYang} in $\chi^{od}$. Figure ~\ref{fig:scaling}(a) shows $\chi^{\rm od}_{R_{x,{\rm max}},\omega =0}$ for different system sizes, where $R_{x,{\rm max}} = (L-1)/2$ is the maximum distance in the $x$ direction on the lattice with periodic boundary conditions. We also plot the size dependence of the cooperon Green's function $G^{b}_{R_{x,{\rm max}},\omega=0}$ in Fig.~\ref{fig:scaling}(b), which behaves very similarly to $\chi^{\rm od}_{R_{x,{\rm max}},\omega =0}$. Both grow substantially with decreasing the temperature, indicating that the Cooper pairs and cooperons become coherent.

\begin{figure}[tbh]
\centerline
{
\includegraphics[width = 5.5cm, height =9cm, angle= 270] 
{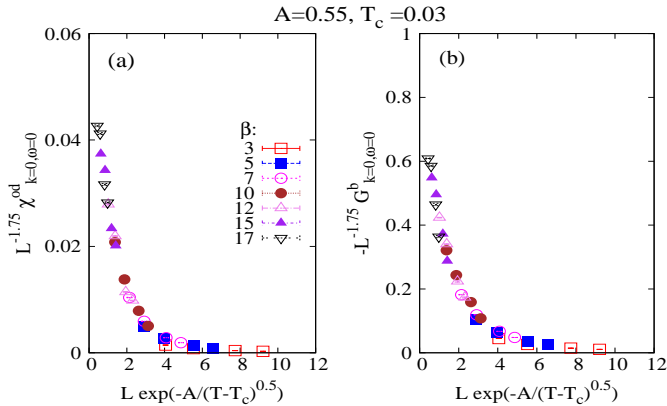} 
}
 \caption[]
 {(Color online.) Finite-size scaling of  $\chi^{\rm od}_{\boldsymbol{k}=0, \omega=0}$ and $G^{b}_{\boldsymbol{k}=0, \omega=0}$ according to Eq.~(\ref{eq:scaling}); the error bars are smaller than the symbol size.
 The uncertainties of $A=0.55\pm0.1$ and $T_{c}=0.03\pm0.01$ are estimated from a breakdown of the shown data collapse.}
\label{fig:GBk0w0-scale}
\end{figure}

We next perform a finite-size scaling analysis for the pair correlator $\chi^{\rm od}_{\boldsymbol{k}=0, \omega=0}$ and the cooperon Green's function $G^{b}_{\boldsymbol{k}=0, \omega=0}$. For $T_{c} > T > 0$, one expects $\chi^{\rm od}_{r,\omega=0}$ to exhibit a power-law decay with an exponent $\eta(T)$, such that $\eta(T_{c}) = 1/4$ and $\eta(T=0)=0$ indicating the emergence of the true log-range order at $T=0$. Above $T_{c}$, $\chi^{\rm od}_{r,\omega=0}$ shows an exponential decay. The pair correlator at $T>T_c$ should obey the scaling formula \cite{Scalapino-PRB-91}
\begin{equation}
\chi^{\rm od} = L^{2-\eta(T_{c})} f(L/\xi^{f}) \mbox{, for $L \gg 1, T \to T^{+}_{c}$}
\label{eq:scaling}
\end{equation}
with $\xi^{f} \sim e^{A/\sqrt{T-T_{c}}}$. 
Since the critical behavior in both subsystems of cooperons and fermions is a manifestation of one and the same superfluid transition, $G^{b}_{\boldsymbol{k}=0, \omega =0}$ is also supposed to exhibit the scaling given by Eq.~\eqref{eq:scaling}. The parameters $A$ and $T_c$ are chosen so that measurements of 
$\chi^{\rm od}_{\boldsymbol{k}=0, \omega =0}$ and $G^{b}_{\boldsymbol{k}=0, \omega =0}$ for different system sizes collapse in the vicinity of the phase transition as shown in Figs.~\ref{fig:GBk0w0-scale}, resulting in $A=0.55\pm0.1$ and $T_{c}=0.03\pm0.01$. The uncertainties in $A$ and $T_{c}$ are estimated from observing a noticeable distortion of the data from a single smooth curve as the parameters are varied beyond the claimed error bars.  

\begin{figure}[tbh]
\centerline
{
\includegraphics[width = 5.5cm, height =9cm, angle= 270] 
{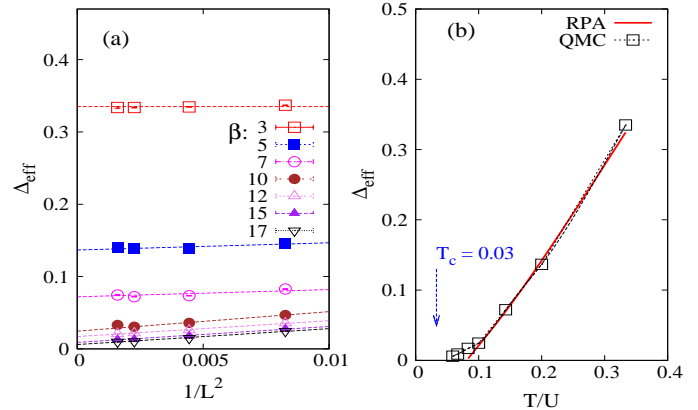} 
}
 \caption[]
 {(a) (Color online.) Finite size extrapolation of the renormalized cooperon gap $\Delta_{{\rm eff}}$. (b) Comparison of the extrapolated $\Delta_{{\rm eff}}$ at $L \to \infty$ with the RPA results. $T_{c} = 0.03$ obtained in Fig.~\ref{fig:GBk0w0-scale} is shown by the arrow. The error bars are smaller than symbol sizes.}
\label{fig:scale}
\end{figure}

The most important properties of cooperons are their effective band gap $\Delta_{\rm eff}$ and effective mass $m_{{\rm eff}}$ renormalized by interactions mediated by fermions. As we shall discuss in more details in the next section, these parameters will allow us to obtain an estimate of the diamagnetic susceptibility, which is expected to rise dramatically due to the quasi-condensation on approach to $T_c$. 
Figure~\ref{fig:scale}(a) shows the dependence of $\Delta_{\rm eff}$, which is obtained from the cooperon Green's function according to $\Delta_{\rm eff} = - [G^{b}_{\boldsymbol{k}=0,\omega=0}]^{-1}$, on the linear system size.
The value $\Delta_{{\rm eff}, L\to \infty}$ in the thermodynamic limit, obtained from the extrapolation in the system size, is shown in Fig.~\ref{fig:scale}(b). At $T>0.1$, the gap  $\Delta_{{\rm eff},L \to \infty}$ is quite close to the RPA estimate. Due to the logarithmic divergence of the bare particle-particle bubble, RPA leads to a substantially higher mean-field critical temperature $T^{MF}_{c} \sim 0.09$.

\begin{figure}[tbh]
\centerline
{
\includegraphics[width = 5.5cm, height =9.0cm, angle= 270] 
{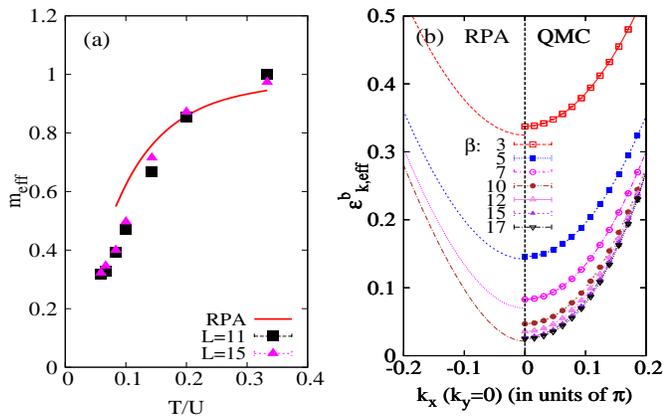}
}
 \caption[]
{
(Color online.) (a):  The suppression of the renormalized cooperon mass with decreasing the temperature. Red curve corresponds to the RPA result. The blue (black) dots are for lattice size $L = 11, 15$. The same fitting process fails for larger lattice sizes $L = 21, 25$ due to the too small measured $G^{b}_{r,\tau}$ close to the lattice boundary and very large sampling error bars.
(b): 
Comparison of the renormalized cooperon's dispersion $\epsilon^{b}_{\boldsymbol{k},eff}$ from RPA calculation and our simulation with $L=11$.
The fitting process (Eq.~\eqref{eq:fit}) is applied to obtain $\epsilon^{b}_{\boldsymbol{k},eff}$ in our simulation.
 The error bars are smaller than the symbol. 
}
\label{fig:dispersion}
\end{figure}

The effective cooperon mass is calculated as
\begin{eqnarray}
m^{-1}_{{\rm eff}} &=& \frac{\sum_{r}G^{b}_{r, \omega =0} r^{2} } {[G^{b}_{\boldsymbol{k}=0, \omega=0}]^{2}},
\label{eq:mass}
\end{eqnarray}
where the factor of $r^{2}$ in the sum shows the importance  of the long-range behaviour of the Green's function in determining $m_{{\rm eff}}$. Since the finite size of the system along with the periodic boundary conditions will enhance $G^{b}_{r, \tau}$ at large $r$, the straightforward evaluation of $m_{{\rm eff}}$ using Eq.~\eqref{eq:mass} is inadequate.
To get rid of the finite-size effects, we fit the measured $G^{b}_{r,\tau}$ according to 
\begin{eqnarray}
G^{b}_{r,\tau} = \sum_{m,n} \mathcal{G}^{b}_{(r_{x} +mL, r_{y}+nL), \tau}
\label{eq:fit}
\end{eqnarray}
with $\mathcal{G}^{b}_{r,\tau} = a e^{-b |r|}$ for large $|r|$ (both $a$ and $b$ depend on $\tau$). The function $\mathcal{G}^{b}_{r,\tau}$ obtained thereby is then used instead of $G^{b}_{r,\tau}$ in Eq.~\eqref{eq:mass}.
 The result obtained using the data for $L=11, 15$ is shown in Fig.~\ref{fig:dispersion}(b). The RPA calculation gives similar values at $T >0.15$. At lower temperatures, the effective mass continuously decreases and tends to a finite value. Fits to Eq.~\eqref{eq:mass} allow us to estimate the renormalized dispersion of the cooperon band $\epsilon^{b}_{\boldsymbol{k},eff}$ as shown in Fig.~\ref{fig:dispersion} using the Green's functions for the case $L=11$. For comparison, the RPA curve $\epsilon^{b}_{\boldsymbol{k},eff}$ is also shown in Fig.~\ref{fig:dispersion}, which agrees with the Monte Carlo results at high temperatures.  

\section{Application to the strong diamagnetism in the cuprates}
Lots of anomalous properties of the pseudogap phase have been reported since the early stage of high-$T_{c}$ studies, for instance the existence of the partial gap, the linear resistivity, and the proportionality of the charge carrier density to doping concentration.\cite{RYZ-Review, pseudogap-review}
Strong superconducting fluctuations have been observed in a large temperature region in recent Nernst and torque magnetometry measurements.\cite{Lu, yayu} In contrast to conventional BCS superconductors, where the Gaussian (amplitude) fluctuations are dominant and the pairing length is quite long, in the cuprates the fluctuations of the phase rigidity are predominant while the cooper pairs are strongly bound with the energy scale around the spin-spin superexchange $J\sim 100$ meV and they localized in a small spatial area with $\xi \sim 3-4$ lattice constant  (a = 3.8 \AA).

In momentum space the pseudogap phase is highly anisotropic. More and more evidence shows that the states at the antinode and node are intrinsically different \cite{Hufner-RPP-08}. 
In the pseudogap phase, the single particle gap is partially opened only around the antinode, leaving either ``arcs" or hole-like Fermi pockets around the nodes.\cite{Norman-98, yang-nature-07,XJZhou-09} Recent angular-resolved photoemission spectroscopy (ARPES) experiments have observed the existence of such pocket and unmask the particle-hole symmetry of the spectrum around the antinode, and the particle-hole asymmetry around the node.\cite{yang-nature-07} Evidence from scanning tunneling microscopy (STM) reveals that the low-energy states around the nodes are homogenous and of long range correlation length, and quantum interference STM observed well-defined Fermi surface only inside AF reduce Brillouin  zone \cite{Kohsaka-nature-08}. Meanwhile the high-energy states around the antinodes are inhomogeneous with short correlation length around four lattice constants. 
Some evidence of particle-hole symmetry in ARPES \cite{yang-nature-07} around the antinodes is consistent with the functional renormalization group calculations,\cite{Honerkamp-RG} showing that the strong umklapp scattering enhances the cooperon channel. 

Some possible scenarios leading to superconductivity by the finite energy cooperon excitations and fermion sea have been proposed from the Hubbard mode on a ladder \cite{KRT10, LeHur-Review} and semiconductors,\cite{Nozieres-EPJB-99} which may shed some light on the case for cuprates.\cite{RYZ-Review} So far the possible interplay between the states residing on the node and the antinode is still an open question. The superconducting gap on the Fermi surface around the node in SC state may be driven by this effect.

\begin{figure}[tbh]
\includegraphics[width=5.5cm,height=6cm, angle=0]
{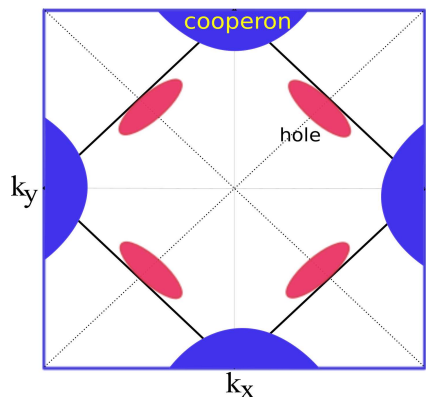}
\caption{(Color online) Anisotropy in momentum space in the pseudogap phase: localized tightly bound cooperons are located at the antinodes and a hole-like  Fermi sea resides on the nodes. 
}
\label{fig:BF1}
\end{figure}

\begin{figure}[b]
\centering
\includegraphics[width=5.5cm,height=7.0cm, angle=0]
{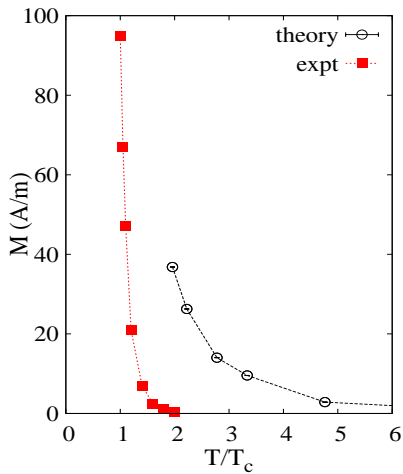}
\caption{(Color online.) The temperature dependence of the magnetization at $B=1$ T calculated using  Eq.~\eqref{eq:chi_b}.The $x$-axis is normalized by $T_{c}\sim 0.03$ ($\sim80$ K with $U=250$ meV). Red dots are from the torque magnetization experiments with $T_c=50{\rm K}$.\cite{Lu} Note that there is no parametric  fitting to the experimental data. While strong damping of cooperons at high temperature suppresses the magnetization at $T \gg T_{c}$, the exponential increase close to $T_{c}$ will not be affected.}
\label{fig:BF}
\end{figure}

It is reasonable to treat the states around the node as a free Fermi gas and the component around the antinode as tightly bound cooperons with a pairing gap of the order of $J\sim 100$ meV as sketched in Fig~\ref{fig:BF1}. 
Our model \eqref{eq:H} can thus be a simplified picture for the cuprate pseudogap phase by ignoring the multi-patch structure and the $d$-wave phase of the momentum-space localized cooperons. There has been a lot of previous work attempting to use this model for the phenomenology of the pseudogap phase (for a review see Ref. \onlinecite{Chen-PR-05}). However most of these are based on mean field calculations, but an exact solution is still missing. Our numerical simulations can give useful quantitative insight. Here we will focus on the strong diamagnetism observed recently. From the recent phenomenological YRZ theory,\cite{YRZ} it is reasonable to further propose that the fermionic particles around the node are itinerant holes, and the cooperons around the antinode are tightly bound hole pairs. To describe the underdoped phase we assume the total charge carrier density is around 0.12, which is obtained by a value of $\mu \sim 100$meV (with $U \sim 250$meV), which is in a reasonable regime based on early analysis of the YRZ model \cite{yang-epl}. The bare hopping $t^{f}= 250$ meV is chosen comparable to the nearest neighbor hopping integral in cuprates.

In conventional BCS theory \cite{Tinkham} the contributions of the fermionic pairing fluctuation to diamagnetism are substantial only in a very narrow temperature region above $T_{c}$. In our model, however, the renormalized cooperon band contributes dominantly to the diamagnetism in a wide temperature range above $T_{c}$ in pseudogap phase. Since it is computationally extremely demanding to calculate the numeric value of the second order coefficient in $\boldsymbol{q}$ of the current-current correlation function $\mathcal{K}_{\boldsymbol{q},\omega=0}$, we approximate the diamagnetization of cooperons as that of free bosons with renormalized gap $\Delta_{\rm eff}$ in the limit $L\to \infty$ and the effective mass $m_{\rm eff}$ at $L=11$ (see Fig.\ref{fig:dispersion}). In  
CGS units [$1/(4\pi)$] it has the form:\cite{May-PR-1959}
\begin{eqnarray}
\chi &=& \frac{(2e)^{2}} {c^{2}m_{{\rm eff}}d} \frac{n^{b}_{\boldsymbol{k}=0}} {6} \label{eq:chi_b},
\end{eqnarray}
where $d$=6$\AA$ is the interlayer distance for cuprates.  The value $(2e)^{2}/(c^{2} 2m_{e}d)$ corresponds to $M$=$7.5 $ A/m at $H=1$ T.

 The experimentally observed singular behavior $M(T,H) \sim -H^{1/\delta(T)}$ with $\delta \to 0$ at small $H$ and $T \to T_{c}^{+}$ might be related to the mesoscopic Meissner effect or the fragile Landon rigidity.\cite{Lu} Thermally excited vortices with exponentially increased inter-vortex length, however, are not sufficient to explain the experimental observation.
In Fig.~\ref{fig:BF} we show the diamagnetism at $H = 1$T with $M = \chi H$ and $\chi$ calculated from Eq.~\eqref{eq:chi_b}. Strong diamagnetism prevails in a very wide temperature region above $T_{c}$; this is in agreement with the experimental data (with $T_{c} = 50 K$). Note that even though the experimental data shows a much narrower temperature region, it is still orders of magnitude wider than that predicted by the conventional BCS theory. In the cooperon-fermion model, the strong damping of the cooperon at high temperature will suppress the diamagnetism greatly but will leave the exponential increase of diamagnetism at $T$ close to $T_{c}$ unchanged. We also note that an alternative explanation of the strong diamagnetism based on a vortex liquid picture  has been proposed by Oganesyan \emph{et al.}. \cite{Oganesyan}

\section{Summary}
We have developed a continuous-time diagrammatic determinant quantum Monte Carlo algorithm for the cooperon-fermion model. 
Our results for the fermionic part of the model show similar behavior to its twin model, the attractive Hubbard model, which is often used to describe the BCS-BEC crossover in the systems of ultra-cold atoms, where the cooperon-fermion model is the relevant model on the BEC side. 

Besides the critical temperature we have calculated the renormalized band gap and mass of the cooperons. The decrease of the mass and the suppression of the renormalized gap have important effects on the thermodynamic properties of the cooperons. Applied to cuprate superconductors, the interplay between the cooperons at the antinode and the fermions at the node is expected to delocalize the cooperons and finally lead to a substantial enhancement of the diamagnetism in a wide temperature range. That could explain the strong diamagnetic signal observed recently in the underdoped state.

The numerical method developed here can be used to study the BCS-BEC crossover on lattices in the framework of the cooperon-fermion model, which gives direct access to the paring physics via the cooperon part. The universal results in terms of to the s-wave effective coupling between the fermions can in principle be obtained in the low-density limit, as was done, e.g., in Ref.~\cite{Burovski-NJP-06}.

We thank T. M. Rice, E. Burovski, M. Sigrist, B. Surer, E. Gull, P. N. Ma., L. Pollet, S. Pilati for discussions. We acknowledge financial support from the Swiss National Science Foundation and the NCCR MaNEP. Xin Wang acknowledge support from the Condensed Matter Theory Center of the University of Maryland.
We used the Brutus cluster at ETH Zurich for most of the simulations.


\begin{thebibliography}{99}

\bibitem{Chen-PR-05} Qijin Chen, J. Stajicb, S. Tanb, K. Levin, Phys. Report \textbf{412}, 1 (2005).
\bibitem{Holland-PRL-01} M. Holland, S. Kokkelmans, M. L. Chiofalo, and R. Walser, Phys. Rev. Lett. \textbf{87}, 120406 (2001) .
\bibitem{Ohashi-PRL-02} Y. Ohashi and A. Griffin, Phys. Rev. Lett. \textbf{89}, 130402 (2002) 

\bibitem{Minca-RMP-90} R. Micnas, J. Ranninger, and S. Robaszkiewicz, Rev. Mod. Phys. \textbf{62}, 113 (1990)
\bibitem{Friedberg-PRB-1989} R. Friedberg and T. D. Lee, Phys. Rev. B \textbf{40}, 6745 (1989).
\bibitem{Domanski-arxiv-03} T. Domanski and J. Ranninger, Phys. Rev. B \textbf{70}, 184503 (2004).
\bibitem{Dima-97} V. B. Geshkenbein, L. B. Ioffe, and A. I. Larkin, Phys. Rev. B \textbf{55}, 3173 (1997).
\bibitem{Micnas-PRB-07} R. Micnas Phys. Rev. B \textbf{76}, 184507 (2007).
\bibitem{Ranning-PC-95} J. Ranning and J. M. Robin, Physica C \textbf{254}, 279 (1995).
\bibitem{Neto-PRB-01} A. H. Castro Neto, Phys. Rev. B \textbf{64}, 104509 (2001).
\bibitem{Perali-PRB-00} A. Perali, C. Castellani, C. Di Castro, M. Grilli, E. Piegari, and A. A. Varlamov, Phys. Rev. B \textbf{62}, R9295 (2000).


\bibitem{Altman-PRB-02}E. Altman and A. Auerbach, Phys. Rev. B \textbf{65}, 104508 (2002).
\bibitem{Tsai-PRB-06} W.-F. Tsai and S. A. Kivelson, Phys. Rev. B \textbf{73}, 214510 (2006).
\bibitem{Alexandrov-04} A. S. Alexandrov, European Physical Journal B, \textbf{39}, 55 (2004).
\bibitem{Robaszkiewicz-PRB-87} S. Robaszkiewicz, R. Micnas, and J. Ranninger, Phys. Rev. B \textbf{36}, 180 (1987).


\bibitem{Nozieres-EPJB-99} P. Nozieres and F. Pistolesi, Eur. Phys. J. B \textbf{10}, 649 (1999). 
\bibitem{KRT10} R. Konik, T. M. Rice and A. M. Tsvelik, Phys. Rev. B \textbf{82}, 054501 (2010). 
\bibitem{LeHur-Review} K. LeHur and T. M. Rice, Ann. Phys. \textbf{324}, 1452 (2009).
\bibitem{YRZ} Kai-Yu Yang, T. M. Rice, and Fu-Chun Zhang Phys. Rev. B \textbf{73}, 174501 (2006).
\bibitem{RYZ-Review} T. M. Rice, Kai-Yu Yang, and Fu-Chun Zhang (unpublished).


\bibitem{ED-BF}M. Cuoco, C. Noce, J. Ranninger, A. Romano, Phys. Rev. B \textbf{67}, 224504 (2003).

\bibitem{sign-problem}K. Binder and D. P. Landau  \textit{A Guide to Monte Carlo Simulations in Statistical Physics} (Cambridge:
Cambridge University Press), 2000;  M. Troyer and U-J Wiese,  Phys. Rev. Lett. \textbf{94}, 170201 (2005).
\bibitem{Burovski-NJP-06} E. Burovski, N. Prokof'ev, B. Svistunov, and M. Troyer, New J Phys. 8 \textbf{153} (2006).
\bibitem{Burovski-PRL-08} E. Burovski, E. Kozik, N. Prokof�ev, B. Svistunov, and M. Troyer, Phys. Rev. Lett \textbf{101}, 090402 (2008).

\bibitem{Lu-epl} Lu Li, Yayu Wang, M. J. Naughton, S. Ono, Yoichi Ando, and N. P. Ong, EuroPhys. Lett. \textbf{72}, 451 (2005).

\bibitem{Schollwock-prl-01} M. Keller, W. Metzner, and U. Schollwock, Phys. Rev. Lett. \textbf{86}, 4612 (2001).


\bibitem{Singer-PRB-96} J. Singer,  M. H. Pedersen, T. Schneider, H. Beck, H.-G. Matuttis, Phys. Rev. B \textbf{54}, 1286 (1996).

\bibitem{CNYang} C. N. Yang, Rev. Mod. Phys. \textbf{34}, 694 (1962).

\bibitem{Scalapino-PRB-91} A. Moreo and D. J. Scalapino, Phys. Rev. Lett. \textbf{66}, 946 (1991)



\bibitem{KT} J. M. Kosterlitz and D. J. Thouless, J .Phys. C: Solid State Phys. \textbf{6}, 1181 (1973); J. M. Kosterlitz,  .Phys. C: Solid State Phys. \textbf{7}, 1046 (1974).



\bibitem{pseudogap-review} P. A. Lee, N. Nagaosa, and Xiao-Gang Wen, Rev. Mod. Phys. \textbf{78}, 17 (2006).

\bibitem{Lu} Lu Li, J. G. Checkelsky, S. Komiya, Y. Ando  and N. P. Ong, Nature physics \textbf{3} 311 (2007); Lu Li, Yayu Wang, S. Komiya, S. Ono, Y. Ando, G. D. Gu, and N. P. Ong, Phys. Rev. B \textbf{81}, 054510 (2010).

\bibitem{yayu} Yayu wang, Z. A. Xu, T. Kakeshita, S. Uchida, S. Ono, Y. Ando, and N. P. Ong, Phys. Rev. B \textbf{64}, 224519 (2001).

\bibitem{Hufner-RPP-08} S. Hufner, M. A. Hossain, A. Damascelli and G. A. Sawatzky, Rep. Prog. Phys. \textbf{71}, 062501 (2008).

\bibitem{Norman-98}M. R. Norman, H. Ding, M. Randeria, J. C. Campuzano, T. Yokoya, T. Takeuchi, T. Takahashi, T. Mochiku, K. Kadowaki, P. Guptasarma  and  D. G. Hinks, Nature \textbf{392}, 157 (1998).

\bibitem{yang-nature-07} H.-B Yang, J. D. Rameau, P. D. Johnson, T. Valla, A. Tsvelik, and G. D. Gu, Nature \textbf{456}, 77 (2008).

\bibitem{XJZhou-09}  J. Meng, Guodong Liu, Wentao Zhang, Lin Zhao, Haiyun Liu, Xiaowen Jia, Daixiang Mu, Shanyu Liu, Xiaoli Dong, Jun Zhang, Wei Lu, Guiling Wang, Yong Zhou, Yong Zhu, Xiaoyang Wang, Zuyan Xu, Chuangtian Chen  and X. J. Zhou, Nature \textbf{462}, 335 (2009)

\bibitem{Kohsaka-nature-08} Y. Kohsaka, C. Taylor, P. Wahl, A. Schmidt, Jhinhwan Lee, K. Fujita, J. W. Alldredge, K. McElroy, Jinho Lee, H. Eisaki, S. Uchida, D.-H. Lee and J. C. Davis, Nature \textbf{454}, 1072 (2008).

\bibitem{Honerkamp-RG} C. Honerkamp, M. Salmhofer, N. Furukawa, and T. M. Rice, Phys. Rev. B \textbf{63}, 035109 (2001).

\bibitem{yang-epl} Kai-Yu Yang, H.-B. Yang, P. D. Johnson, T. M. Rice, Fu-Chun Zhang, Eupo. Phys. Lett. \textbf{86}, 37002 (2009).

\bibitem{Tinkham} M. Thinkham, \textit{ Introduction to Superconductivity (2ed. McGraw-Hill, 1996)}. 

\bibitem{May-PR-1959} R. M. May, Phys. Rev. \textbf{115}, 254 (1959)

\bibitem{Oganesyan} V. Oganesyan, David A. Huse, and S. L. Sondhi, Phys. Rev. B \textbf{73}, 094503 (2006)

\end{thebibliography}
\end{document}